%Paper: astro-ph/9308010
%From: falchi@arcetri.astro.it (Ambretta Falchi)
%Date: Fri, 6 Aug 1993 14:54:28 +0200

\vsize  22truecm
\hoffset=-0.5truecm\voffset=0.3truecm
\nopagenumbers\parindent=0pt
\footline={\ifnum\pageno<1 \hss\thinspace\hss
    \else\hss\folio\hss \fi}
\pageno=-1

\newdimen\windowhsize \windowhsize=13.1truecm
\newdimen\windowvsize \windowvsize=6.6truecm

\def\heading#1{
    \vskip0pt plus6\baselineskip\penalty-250\vskip0pt plus-6\baselineskip
    \vskip2\baselineskip\vskip 0pt plus 3pt minus 3pt
    \centerline{\bf#1}
    \global\count11=0\nobreak\vskip\baselineskip}
\count10=0
\def\section#1{
    \vskip0pt plus6\baselineskip\penalty-250\vskip0pt plus-6\baselineskip
    \vskip2\baselineskip plus 3pt minus 3pt
    \global\advance\count10 by 1
    \centerline{\expandafter{\number\count10}.\ \bf{#1}}
    \global\count11=0\nobreak\vskip\baselineskip}
\def\subsection#1{
    \vskip0pt plus3\baselineskip\penalty-200\vskip0pt plus-3\baselineskip
    \vskip1\baselineskip plus 3pt minus 3pt
    \global\advance\count11 by 1
    \centerline{{\it {\number\count10}.{\number\count11}\/})\ \it #1}}
\def\firstsubsection#1{
    \vskip0pt plus3\baselineskip\penalty-200\vskip0pt plus-3\baselineskip
    \vskip 0pt plus 3pt minus 3pt
    \global\advance\count11 by 1
    \centerline{{\it {\number\count10}.{\number\count11}\/})\ \it #1}}
\def\etal{{\it et al.\/}}
\def\eol{\hfil\break}
\def\affl#1{\noindent\llap{$^{#1}$}}

%TITOLO
{
\def\cl#1{\hbox to \windowhsize{\hfill#1\hfill}}
\hbox to\hsize{\hfill\hbox{\vbox to\windowvsize{\vfill
\bf
\cl{ ATMOSPHERIC MODELS OF FLARE STARS:}
\cl{ THE QUIESCENT STATE OF AD LEO }
\bigskip
\cl{Pablo J.D. Mauas$^{1,2}$ and Ambretta Falchi$^{1}$}
\bigskip\rm
\cl{Preprint n.~29/93}

\vfill}}\hfill}}

%AFFILIAZIONI
\vskip5truecm
{\leftskip1.7truecm
\affl{1}Osservatorio Astrofisico di Arcetri,
\eol
Largo E.~Fermi 5, I-50125 Firenze (Italy)
\bigskip
\affl{2}Instituto de Astronom\'\i a y F\'\i sica del Espacio,
\eol
CC 67, Suc. 28, 1428 - Buenos Aires, R. Argentina

\vfill
Astronomy and Astrophysics, in press
\vglue3truecm
}
\eject

%ABSTRACT
\vglue\windowvsize plus 0pt minus \vsize
\heading{ABSTRACT}

        We compute a semi-empirical atmospheric model for the dMe star
        AD Leo, which constitutes the first model computed to match the
        continuum observations, as well as a wide set of chromospheric spectral
        lines. We find good agreement between the computed and observed
spectral
        features, with the exception of the Ca II K line.

\medskip
\noindent\underbar{\strut Subject}\ \underbar{\strut headings}:
{Stars: atmospheres ; Stars: chromosphere ;Stars: late-type ; Stars: Flare }

\vglue\windowvsize plus 1fill minus \vsize
\eject

%TEXT
\pageno=1
\tracingmacros=1
\tracingcommands=2
\def\matono#1 {\ifmmode #1 \else $#1$\fi}
\def\sub#1 {\matono _{\rm #1} \ifspace}
\def\u#1 {\matono ^{\rm #1} \ifspace}
\def\SUB#1{$_{\rm #1}$}

\def\eqnu#1 {\eqno{(\rm{#1})}$$}
\sfcode`\ =1001 \sfcode`"=1001
\def\ifspace#1{\ifmmode #1\else \ifnum\the\sfcode`#1 >
	1000 #1\else \ifnum\the\sfcode`#1 = 0 #1\else
	{{ }#1}\fi \fi \fi}
\newif\ifcoma
\def\sinumero#1{\ifnum12=\the\catcode`#1 \bf#1\comatrue\else #1\comafalse\fi}
%----------------------
\def\Avr{Avrett E.H.\ifspace}

\def\Kur{Kurucz R.L.\ifspace}
\def\Loe{Loeser R.\ifspace}
\def\Mau{Mauas P.J.\ifspace}

\def\MAL{Mauas, Avrett, \& Loeser\ifspace}
%----------------------
\def\ie{\hbox{\it i.e.}\ifspace}
\def\eg{\hbox{\it e.g.}\ifspace}
\def\lin#1 {\matono{\lambda } #1~\AA \ifspace}
\def\La{Ly{\matono_{\alpha} }\ifspace}
\def\Ha{H{\matono_{\alpha} }\ifspace}
\def\Hb{H{\matono_{\beta} }\ifspace}
\def\Hg{H{\matono_{\gamma} }\ifspace}
\def\Hd{H{\matono_{\delta} }\ifspace}
\def\Tm{T{\matono_{min} }\ifspace}

\def\Hm{H{\matono^- }\ifspace}

\def\WSG{Worden et al.\ifspace}
\def\PC{Pettersen \& Coleman\ifspace}
\def\GLSW{Giampapa et al.\ifspace}
\def\n{\sub{\nu } }
\def\m{$\mu$m}
%------------------------------
\def\hf{\hfil}
\def\cuad{\hskip 7pt\relax}    % 0.7 x \quad
\def\E{E$-$}

\sfcode`.=0 \sfcode`,=0 \sfcode`)=0

\hoffset=-0.5truecm\voffset=0.3truecm
\nopagenumbers\parindent=0pt
\footline={\ifnum\pageno<1 \hss\thinspace\hss
    \else\hss\folio\hss \fi}
\pageno=1

\section{ Introduction}

Dwarf M stars are divided in two subclasses according to whether
the Balmer lines are in emission or not. This fact distinguishes
the emission dMe stars from the non-emission dM stars. The presence, in both
types, of many atomic lines and molecular bandheads in absorption, proves the
existence of a cool photosphere. Emission lines are instead signatures
of a chromosphere with a considerable temperature rise.

The difference between the spectra of dM and dMe stars could be similar to
that between quiet and active regions on the Sun. The possible heating
mechanism of the chromosphere is supposed to be more efficient in
correspondence
to a concentration of the magnetic field, as in solar faculae.  The presence on
dMe stars of strong magnetic fields, with an area coverage greater than 70 \%
(Saar 1990), supports the idea that most of the stellar surface is covered
by facular network and that the contribution to the emitted spectrum comes
essentially from these bright regions.

Atmospheric models for a grid of dM stars were computed by
Mould (1976), from the deep photosphere to approximately the
temperature minimum. The temperature minimum region and the
chromosphere are instead poorly modeled.  Cram \& Mullan (1979) computed  very
schematic chromospheric models using only Balmer-line data as chromospheric
diagnostics; Giampapa et al. (1982) computed
single component homogeneous models that are consistent with high resolution
profiles of the Ca II K line for some representative dMe and dM
stars. Their models have microturbulent velocities between 1 and 2 km s$^{-1}$.

To study the atmospheric properties of a dMe star we chose
AD Leo (Gl 388), which is one of the flare stars most extensively observed
both in the quiescent and in the flaring states. It is a single red dwarf star,
on the main sequence, with an effective temperature of 3500 K (Pettersen 1980),
and a spectral type M3.5e (Kunkel 1975). Vogt
et al. (1983) measured a projected rotational velocity $v
\sin(i)$ of 5 km s$^{-1}$ from an absorption line profile (Ba II
\lin{6141} ) at high resolution, and suggested that such a rapid rotation is
the
necessary condition for the existence of photometric modulation by
starspots and active regions
(the BY Draconis syndrome). This possibility was confirmed by Spiesman
\& Hawley (1986), who measured a photometric period P = $2^d.7 \pm .05$.

High resolution spectral observations % outside of flares,
of the quiescent state	show chromospheric emission lines that
range from a weak emission core in the center of the Na I D (5891,
5896 \AA) and the Ca II IR lines (8498, 8542, 8668 \AA) to a strong
emission in the He I D$_3$, Mg II h and k,
Ca II H and K and H I Balmer lines (Giampapa et al. 1978, Pettersen \&
Coleman 1981, Worden et al. 1981, Doyle 1987). Sundland et al. (1988) found
that the Balmer lines are the most important components to the chromospheric
radiation loss.

AD Leo is an active flare star: on the average one outburst is detected every
1 - 1.5 hours.  Long term changes have been analyzed for the period between
1972 - 1988 (Pettersen et al. 1984, 1986, 1990) and only marginal variations in
the flare activity have been found.

Hawley \& Fisher (1992) computed a theoretical model of the quiescent
atmosphere
of AD Leo from the corona to the photosphere.  The structure of the corona
and the transition region is computed for a loop of length L = 10$^{10}$ cm
and an apex coronal temperature of $3 \times 10^6$ K, in hydrostatic and
energetic equilibrium. The photosphere is taken from the grid of models
by Mould (1976). The chromosphere is simply an
interpolation joining the transition region and the photosphere. They
consider this atmospheric model only as a schematic one, chosen to test a
certain heating model of the flare atmosphere and not to fit the observed
spectral features of the star.

In this paper we present an atmospheric model of AD Leo, computed to fit
several spectral features. In \S 2 we present our model, and discuss the
fit to the observations. In \S 3 we compute the radiative losses to be
balanced by the chromospheric heating. Finally, in \S 4 we discuss the
results.

\section{ The atmospheric model}

We compute a semi-empirical model of the photosphere and
chromosphere of AD Leo. For semi-empirical we mean that the model was
computed to match a certain number of spectral features as well as
possible. For a given model, we compute the continuum fluxes, the profiles for
the Balmer, Na D, Mg b, and Ca II K and \lin{8499} lines, and the total
fluxes in the \La and the Mg II h and k lines, which are compared
with the observations. The atmospheric model is then iteratively modified to
improve the match.

In this paper we use the observations published by Pettersen \& Hawley
(1987), which have a 3.5 \AA\ resolution. Although for some lines
there are observations with better resolution, we consider that it is
important to take a consistent set of observations, made at the same time
and calibrated uniformly.

The calculations were done using the computer code Pandora, kindly provided by
Dr. E. Avrett (see Avrett \& Loeser 1992 for an explanation of the program's
features). Given a temperature vs. height distribution, the hydrogen
density at each height is computed assuming hydrostatic equilibrium
(HSE). The set of statistical equilibrium equations, coupled with the radiative
transfer equation, is then solved to compute the electron density and the
emergent spectrum. These calculations are done self-consistently for H, He I,
Mg I, Na I, Fe I, Si I, Ca I and Al I (solar abundances are assumed).

We used a microturbulent velocity between 1 and 2 km s\u-1 ,
both in the HSE equation and for the line-profile calculations. Although
this is a rather arbitrary choice, the values are similar to the
ones given by Giampapa et al. (1982).

Observations of magnetically sensitive infrared lines by Saar \& Linsky
(1985), give a value for the filling factor (relation of the facular area
to the total area of the star) of $0.73\pm0.06$. Given the high value of
the filling factor, we follow the standard assumption found in the literature,
and neglect the contribution of the non-facular area to the total brightness
of the star. Thus, we assume that a one component plane-parallel model is a
good approximation to the mean state of the stellar atmosphere.

The model that represents the best fit to the chosen set of observations is
shown in Fig. 1, together with the region of formation of the lines
used in the modeling. For reference, we also show in Fig. 1
the model obtained by Hawley \& Fisher (1992).
The atmospheric parameters are listed in Table 1.
The columns given represent the column mass
in g cm$^{-2}$, the electron temperature in K, the microturbulent velocity
in km s\u-1 ; the continuum optical depth at 5000 \AA; the hydrogen,
proton, and electron density in particles per cm\u3 ; and the height
(in km) above the level where $\tau$\SUB{5000\AA} = 1.

\vfill
\eject
{\it Table 1}Atmospheric parameters for the model discussed in this paper
\bigskip
\vbox{ \tabskip=.5truecm plus1truecm
\halign  to 16truecm
{\hf#\hf & \hf# & \hf#\hf & \hf#\hf & \hf#\hf & \hf#\hf & \hf#\hf & \hf# \cr
       \noalign{\hrule\vskip .07cm\hrule\vskip 8pt}
m&T\hf&V\SUB{t}&$\tau$\SUB{5000\AA}&N\SUB{H}&N\SUB{p}&N\SUB{e}&h\hf\cr
       \noalign{\vskip 3pt}
(g cm\u-2 )&(K)\hf&(km/s)&   &(cm\u-3 )&(cm\u-3 )&(cm\u-3 )&(km)\hf\cr
       \noalign{\vskip 5pt\hrule\vskip 5pt}
5.23\E06 & 145000 &  2.07 & 0.00E+00 & 6.37E+09 & 6.37E+09 & 7.66E+09 & 1125\cr
5.23\E06 & 122300 &  2.07 & 2.22\E09 & 7.56E+09 & 7.56E+09 & 9.09E+09 & 1121\cr
5.24\E06 & 103100 &  2.06 & 4.86\E09 & 8.98E+09 & 8.98E+09 & 1.08E+10 & 1117\cr
5.25\E06 &  86980 &  2.06 & 7.99\E09 & 1.07E+10 & 1.07E+10 & 1.28E+10 & 1113\cr
5.26\E06 &  73360 &  2.06 & 1.17\E08 & 1.27E+10 & 1.27E+10 & 1.52E+10 & 1109\cr
5.27\E06 &  61870 &  2.05 & 1.61\E08 & 1.51E+10 & 1.50E+10 & 1.81E+10 & 1105\cr
5.29\E06 &  50000 &  2.05 & 2.28\E08 & 1.87E+10 & 1.87E+10 & 2.24E+10 & 1100\cr
5.34\E06 &  45590 &  2.04 & 3.86\E08 & 2.07E+10 & 2.07E+10 & 2.48E+10 & 1090\cr
5.42\E06 &  39700 &  2.03 & 6.53\E08 & 2.41E+10 & 2.41E+10 & 2.89E+10 & 1075\cr
5.48\E06 &  36200 &  2.02 & 8.56\E08 & 2.67E+10 & 2.67E+10 & 3.20E+10 & 1065\cr
5.58\E06 &  31250 &  2.00 & 1.20\E07 & 3.16E+10 & 3.15E+10 & 3.76E+10 & 1050\cr
5.81\E06 &  23000 &  1.98 & 1.93\E07 & 4.75E+10 & 4.66E+10 & 4.99E+10 & 1025\cr
6.14\E06 &  18000 &  1.97 & 2.96\E07 & 6.62E+10 & 6.30E+10 & 6.49E+10 & 1000\cr
8.44\E06 &  14000 &  1.97 & 1.24\E06 & 1.18E+11 & 1.12E+11 & 1.14E+11 &  890\cr
1.20\E05 &  12000 &  1.96 & 3.26\E06 & 1.97E+11 & 1.86E+11 & 1.87E+11 &  790\cr
1.60\E05 &  10830 &  1.95 & 5.79\E06 & 2.95E+11 & 2.67E+11 & 2.68E+11 &  720\cr
2.36\E05 &   9499 &  1.94 & 1.07\E05 & 5.65E+11 & 3.72E+11 & 3.73E+11 &  640\cr
3.72\E05 &   8415 &  1.93 & 1.58\E05 & 1.35E+12 & 2.58E+11 & 2.60E+11 &  575\cr
8.43\E05 &   7064 &  1.92 & 1.97\E05 & 4.19E+12 & 5.88E+10 & 6.36E+10 &  494\cr
1.45\E04 &   6330 &  1.90 & 2.08\E05 & 8.11E+12 & 2.75E+10 & 3.31E+10 &  450\cr
2.19\E04 &   5830 &  1.88 & 2.14\E05 & 1.33E+13 & 1.56E+10 & 2.16E+10 &  420\cr
3.38\E04 &   5600 &  1.82 & 2.21\E05 & 2.13E+13 & 1.21E+10 & 2.08E+10 &  390\cr
5.31\E04 &   5400 &  1.76 & 2.31\E05 & 3.49E+13 & 1.28E+10 & 2.56E+10 &  360\cr
8.50\E04 &   5200 &  1.72 & 2.49\E05 & 5.80E+13 & 1.97E+10 & 3.74E+10 &  330\cr
1.39\E03 &   5000 &  1.67 & 2.88\E05 & 9.84E+13 & 4.14E+10 & 6.20E+10 &  300\cr
2.31\E03 &   4800 &  1.65 & 3.63\E05 & 1.70E+14 & 5.53E+10 & 8.01E+10 &  270\cr
3.62\E03 &   4400 &  1.61 & 4.48\E05 & 2.91E+14 & 3.32E+10 & 7.04E+10 &  245\cr
5.94\E03 &   3956 &  1.59 & 5.76\E05 & 5.30E+14 & 1.85E+10 & 7.97E+10 &  220\cr
9.20\E03 &   3600 &  1.57 & 8.07\E05 & 8.99E+14 & 1.09E+10 & 1.05E+11 &  200\cr
2.35\E02 &   3000 &  1.53 & 3.76\E04 & 2.82E+15 & 1.06E+09 & 2.34E+11 &  163\cr
6.72\E02 &   2780 &  1.48 & 1.83\E03 & 1.03E+16 & 6.28E+06 & 2.61E+11 &  130\cr
2.26\E01 &   2700 &  1.44 & 5.52\E03 & 4.22E+16 & 5.95E+03 & 2.78E+11 &  100\cr
8.70\E01 &   2800 &  1.40 & 1.66\E02 & 1.65E+17 & 1.10E+03 & 4.04E+11 &   70\cr
2.03E+00 &   3014 &  1.37 & 4.25\E02 & 3.47E+17 & 4.76E+03 & 7.15E+11 &   50\cr
5.20E+00 &   3282 &  1.31 & 1.95\E01 & 7.88E+17 & 8.72E+04 & 1.96E+12 &   25\cr
1.20E+01 &   3550 &  1.29 & 1.00E+00 & 1.62E+18 & 3.05E+06 & 5.53E+12 &    0\cr
2.50E+01 &   3900 &  1.23 & 4.75E+00 & 2.89E+18 & 1.58E+08 & 1.53E+13 &  -25\cr
4.60E+01 &   4400 &  1.18 & 2.06E+01 & 4.28E+18 & 1.21E+10 & 4.76E+13 &  -50\cr
7.68E+01 &   4800 &  1.18 & 7.43E+01 & 6.30E+18 & 1.80E+11 & 1.19E+14 &  -75\cr
1.20E+02 &   5300 &  1.18 & 2.30E+02 & 8.66E+18 & 2.76E+12 & 3.00E+14 & -100\cr
1.79E+02 &   5852 &  1.18 & 6.14E+02 & 1.15E+19 & 3.28E+13 & 6.75E+14 & -125\cr
\noalign{\vskip 4pt\hrule\vskip 4pt}
}}

\firstsubsection{The continuum calculations}
In Fig. 2 we compare the observed continuum spectrum (smoothed to
5 \AA) with the results obtained for
this model.
The error of these observations has been estimated at
around 10\% (Hawley, private comunication).

%\begfigwid 16 cm
%\begfigwid 1 cm
%\figure2{Comparison of the observations (full line) in the continuum with the
%results computed for our model, with (dashed line) and whitout (dotted
%line) considering line blanketing. Also indicated are the heads of the most
%important molecular bands}
%\endfig

A very important factor to take into account when modeling a cool
star like AD Leo, is the inclusion in the opacity calculations of the line
blanketing due to the numerous weak atomic lines,
and to the molecular species present in the atmosphere, in particular TiO
and CaOH  (Mould 1976). In this work we included the $5.8 \times 10^6$ atomic
and molecular lines computed by Kurucz (1991a). It should be pointed out
that these calculations do not include any triatomic molecules, that are
thought to be important for this type of stars. The heads of the most important
molecular bands observed in AD Leo are indicated in Fig. 2.

In Fig. 2, the dotted line represents the results obtained for this
atmospheric model, but without considering line blanketing.  It is clear that
the inclusion of this opacity source (dashed line in Fig. 2)
 is of great importance to match the observed
spectrum. It should be pointed out that line blanketing affects not
only the computed continuum spectrum, but also the equilibrium of
the different atomic species and hence the computed model.

As it is seen in Fig. 2, although the general behaviour of the observed
spectrum is reproduced, and many observed spectral features are matched,
there are still some molecular bands clearly absent from our calculations,
notably CaOH and H$_2$O. It can also be noted that the opacity due to TiO
seems to be overestimated. Kurucz (1991b) cautions against the use of his
opacity calculations to compute the atmospheres of M stars. However, based
on the present results, we believe Kurucz's compilation is quite good even
for this case.

In Table 2 we list the observed flux values obtained with the Johnson UBVR
filters and with IR filters, together with the fluxes computed from our
model (integrated with the corresponding filter profile, see Schaifers and
Voigt 1982; for the IR filters, we used a square filter profile).
As it can be seen, the general agreement is good.

Note the large differences for the visible filters
between the integrated fluxes with and without line blanketing. This
implies that models based on a fit to the filter values, computed without
taking into account these opacities, can be strongly in error.

%\begtabfull
{\it Table 2} Observed and computed fluxes at earth (erg cm\u-2 s\u-1 )
for different spectral features.
 For the optical continuum filters, the first value was computed
considering line blanketing, and the second one without considering it

\vbox{ \tabskip=.4truecm plus1truecm
\halign to 8.8truecm
{&\hf\hf#\hf\cr
       \noalign{\hrule\vskip .07cm\hrule\vskip 8pt}
filter & ref & obs &computed\cr
       \noalign{\vskip 3pt}
       \noalign{\vskip 5pt\hrule\vskip 5pt}
U & 1  &  6.5\E14  &  4.2\E14  \cuad  6.2\E13 \cr
B & 1  &  2.9\E13  &  2.7\E13  \cuad  7.2\E13 \cr
V & 1  &  6.6\E13  &  7.1\E13  \cuad  1.4\E12 \cr
R & 1  &  1.5\E12  &  1.1\E12  \cuad  2.0\E12 \cr
       \noalign{\vskip 3pt}
12\m & 2 & 7.1\E24  &  8.8\E24  \cr
25\m & 2 & 2.4\E24  &  1.6\E24  \cr
       \noalign{\vskip 6pt}
\La {} & 3 & 1.0\E12  &  2.9\E12  \cr
Mg II h+k  & 3 & 2.5\E12& 1.1\E12\ +\ 1.4\E12 \cr
Ca II K  & 3 & 1.2\E12 & 2.5\E13 \cr
\Ha {} & 3 & 4.4\E12  &  5.9\E12  \cr
\Hb {} & 3 & 2.4\E12  &  2.1\E12  \cr
\Hg {} & 3 & 1.2\E12  &  1.0\E12  \cr
\Hd {} & 3 & 8.7\E13  &  6.3\E13  \cr
\noalign{\vskip 4pt\hrule\vskip 4pt}
\multispan {3} {}\u1 Pettersen (1976)\hf \cr
\multispan {3} {}\u2 Mathioudakis and Doyle (1991)\hf \cr
\multispan {3} {}\u3 Hawley (1989)\hf \cr
}}
%\endtab

Finally, we would like to point out that the good agreement between the
observed and computed spectra implies that the photospheric part of our
model can be considered quite reliable.

{}

\subsection{The Balmer lines}

Figure 3 compares the computed profiles for the first Balmer lines
with the observations (H\sub{\epsilon } is not shown, because it appears
blended with Ca II H). Since the observations have a spectral resolution
of 3.5 \AA, we have smeared our computed profiles with a gaussian
profile of
3.5 \AA\ FWHM. When comparing the profiles, we plot the computed
profile with a dashed line, and the smeared one with a full line.
In Figure 4 we show the source function and the Planck function for
\Ha and \Hg, together with those for the Na D\sub1 \ and the Ca
II K lines. We also indicate in this figure the depth of formation at
line center for two different values of $\mu$ ($=\cos\theta$).
The computed and observed line fluxes are listed in Table 2.

%\begfigwid 16 cm
%\begfigwid 1 cm
%\figure3{Comparison of the observations (crosses) in the Balmer lines and the
%results computed for our model. Here and in the next figures, the dashed
%line shows the computed profile with ``infinite" spectral resolution, and the
%full line shows the calculations smeared to the same resolution than
%the observations}
%\endfig

%\begfigwid 16 cm
%\begfigwid 1 cm
%\figure4{Planck function ($B_\nu$, full line) and source function
%($S_\nu$, dashed line) for different lines. The depth of formation (peak
%of the contribution function) of the line center is also shown. The
%values at disk center are indicated above the $S_\nu$ curve, and those
%at $\mu=0.6$ are indicated below it}
%\endfig

Some observations with better spectral resolution are
available for \Ha (Giampapa et al. 1978; \WSG (1981);
 Pettersen \& Coleman 1981). We
will attempt a  comparison of our calculations with these
profiles, even if they were taken in different conditions than
the ones modeled here.

Giampapa et al. (1978) and Worden et al. (1981), with spectral resolutions
of the order of 0.25 \AA, measured a contrast (I\sub peak /I\sub continuum )
of 2.05 and 6.57 respectively. Our computed contrast, after convolution
with the appropiate instrumental profile, is 5.77. \PC (1981) measured a
contrast of 3.1, with a spectral resolution of 0.45\AA. Our computed
contrast (after convolution) is 4.9.

In our calculations we find for \Ha a FWHM of 1.4 \AA, which is in good
agreement with the observed values of 1.4, 1.35, and 1.29 \AA\ (Giampapa et
al. 1978; \PC 1981; \WSG 1981, respectively). The separation between the
computed emission peaks is 1. \AA, while the observed one is 0.6 \AA\ (\PC
1981).

It should be
noted that the self reversal of \Ha that can be seen in the computed profile
before convolution with the instrumental response (Figure 3, dotted line),
 can also be
seen in two of the quoted papers. After convolution with the instrumental
profile appropriate to each observation we find a self reversal of the
order of 1.6, which is larger than observed.

The  observed differences may be ascribed
to different levels of chromospheric activity of the star.
The present model, which has been computed to match a given set of
observations, yields an \Ha profile comparable with other observations
within a factor of two, and may be considered a good starting point to
represent an average atmosphere of AD Leo.

\subsection{The Na I lines}

Figure 5 shows the observed and computed profiles for the Na I D lines
and for the infrared lines at \lin{8183} and \lin{8195} .
The source and Planck function for the D\sub1 line are shown in Fig. 4.

%\begfigwid 16 cm
%\begfigwid 1 cm
%\figure5{Comparison of the observations in the Na D and infrared  lines and
%%the
%results computed for our model. See fig. 3}
%\endfig

It can be seen that, while the agreement in the infrared lines is quite good,
the difference between computed and observed profiles for the D lines is
larger. It should be noted, however, that the D lines lie over two
molecular bands, of TiO and H$_2$O.

The D lines were also observed, at better spectral resolution, by \GLSW (1978),
\WSG (1981) and \PC (1981). In all these cases a central emission peak,
unresolved in the Pettersen \& Hawley observations, has been seen.
In our computed spectrum, besides a central emission, we
find a self-reversal that cannot be resolved at the resolution of neither
set of observations.

\subsection{The Mg b lines}

The formation of the Mg I b lines have been successfully applied to solar flare
atmospheric models (see Mauas et al. 1988, 1990; Metcalf et al. 1990).
For this reason, it
seemed natural to choose them as one of the features to be reproduced by
our model. However, in the spectrum of cool stars the b lines lie over the head
of a molecular TiO band (at 5167 \AA), and thus observations at low
resolution do not provide reliable information on their intensity.
In any case, a comparison between the observed and computed
spectra is shown in Fig. 6. Unfortunately, no better resolution
observations of these lines are available.

%\begfigwid 8 cm
%\begfigwid 1 cm
%\figure6{Comparison of the observations in the Mg b lines and the
%results computed for our model. See Fig. 3}
%\endfig

\subsection{The Ca lines}

A line sometimes used for spectral classification is the Ca I \lin4227 .
It seems to be the broadest line in the quiet spectrum of AD Leo,
with a FWHM of 13 \AA\ (see Fig. 7). However, in the flare spectra
at these wavelengths three emission features are visible (dotted lines
in Fig. 7), one corresponding to the central wavelength of the Ca line, and
the other two in the supposed wings.

The more obvious explanation is the presence of other strong metallic
lines, some of which turn into emission during the flare. If this is the
case, other metallic lines are blended with Ca I \lin4227 in such a way
that, at the present spectral resolution, we cannot separate the different
contributions in the quiet spectrum. For this reason, we did not use this
line to construct our model.

%\begfig 9 cm
%\figure7{Comparison of the observations in the Ca I \lin4227 in the quiet
%spectrum (solid line) and the flare spectrum at two different times (dotted
%line)}
%\endfig

Figure 8 shows a comparison between the computed and observed
spectra for the Ca II K and the infrared line at \lin{8498} (we also
computed the lines at \lin{8542 and 8668} which are not shown here).
The source and Planck functions for the K line can be found in Fig. 4,
and its computed and observed line fluxes are compared in
Table 2.
It can be seen that, even if the agreement for the IR line is acceptable,
the computed intensity for the K line is only a fourth of what is observed.

%\begfigwid 8 cm
%\begfigwid 1 cm
%\figure8{Comparison of the observations in the Ca II K and infrared line and
%%the
%results computed for our model. See Fig. 3}
%\endfig

Since the Ca II K line and \Hg have the same height of formation, we used
the relation I(\Hg)/I(Ca II K) as a parameter to test our models. For the
final model, the computed value of this intensity ratio is 3.3
times larger than the observed one, and in the different models we tried
the relation between the computed and observed ratio
varied from 3 to 7. There is no way to diminish this relation without
destroying the agreement for the other lines.

For this reason, we believe
that the problem is not in the atmospheric model, but in the way the lines
are treated. For example,
it is worth noting that, in the present calculations, collisions with
hydrogen atoms were not taken into account in the statistical
equilibrium equation, except for the hydrogen lines.
However, in a cool star like
AD Leo, these collisions can be of importance, bringing the solution
closer to local thermodynamic equilibrium, which for the Ca II K line would
mean
a higher computed intensity. To quantify this effect, we made a
calculation including (for Ca II) collisions with hydrogen assuming a
hydrogenic rate (Kanlakys 1985). The computed central intensity did in
fact increase, but only by a 50\%.
However, the approximation we used was
quite crude, and we believe the difference with the observations can be
due to the effect of collisions with hydrogen atoms.

Another factor that should be considered when computing the Ca II H and
K lines is that these lines are affected by partial
redistribution (PRD). Even if this effect is important mainly in
the line wings, differences in the PRD parameters adopted do alter the core
intensity. For the K line, changing the PRD parameters can change the
central intensity by as much as 20\%. The profiles shown here are
the larger ones we could obtain, which correspond to a broad complete
redistribution core (for details on how PRD is treated, see \MAL 1989).

\subsection{\La and the Mg II h and k lines}

Hawley (1989) gives an integrated flux for the \La
line of 1~10\u-12 erg cm\u2 s$^{-1}$. Our computed values
are too large, and even adjusting the PRD parameters we can only obtain
a value of 2.9 10\u-12 erg cm\u2 s$^{-1}$. It should be noted that, while
the computed profiles for \La are larger than the observations, the
inverse is true in the case of the Ca II K line. Thus,
the agreement in the Ca II K line is best with a very broad
complete redistribution core (as is the case for the Mg II lines, see
below), which increases the total flux in the line, while for \La a large
amount of PRD is needed.

Hawley (1989) gives a total observed line flux for the Mg~II h
and k lines of 2.5 10\u-12 erg cm\u-2 s$^{-1}$. We can reproduce this
value, adjusting the PRD parameters, as can be seen in Table 2.
High resolution IUE observations of these lines were published by Ambruster
et al. (1989). They give a total flux in agreement with Hawley (1989),
an average flux of the h line of 1.13 $\pm$ 0.11 erg cm\u-2 s$^{-1}$ and a
flux
of the k line (not averaged for saturation problems) of 1.4 erg cm\u-2
s$^{-1}$ which are in very good agreement with our computed values.
However, our computed profiles
have a central intensity twice as large as the observed ones, and half
the FWHM. It is worth noting that in the present calculations we did not
include any macroturbulence, which can correct part of this difference (see
e.g. Lites \& Skumanich 1982).

\subsection{Discussion}

Some conclusions can be drawn from
comparison of our model with the one by Hawley \& Fisher (1992, HF). The
first one is that the photospheric part of both models are quite
similar. In fact, the photosphere of the model by HF was
taken from the work by Mould (1976), who constructed a grid of
photospheric models for M dwarfs. As the main opacity source in the continuum
is given by the different molecular species, and in both works these
opacities are treated in LTE, this agreement is not surprising.
However, we should point out that Mould's work included the opacity due
to H$_2$O, while we did not include any triatomic molecules.

On the other hand, the chromospheres of both models are very different.
Regarding this difference, we should note
that HF simply interpolated between their computed, theoretical,
transition region and Mould's photosphere. In our case, the chromosphere
was computed to match a set of observations.

As indicated in
Fig. 1, there is a lack of spectral features originating in the
temperature minimum and the low chromosphere, which could serve as
reliable diagnostics of the structure of this region. On the other hand,
the source function of the chromospheric lines depends
strongly on the structure of the \Tm (see Fig. 4), and therefore these
lines give some indication on this region.
In fact, we needed a broad \Tm to avoid a strong emission in the Mg b
and Na D lines, which is not observed. However, it would be very
helpful for future chromospheric modeling if some lines formed at the
\Tm were available. These lines  are
weaker and narrower  than the chromospheric ones, thus a much
better spectral resolution is needed.

We can also see in Fig. 1 that HF's transition region is
placed at a lower column mass than ours.
This is a consequence of their theoretical
approach, starting from a specific physical model for the corona
that determines the structure of the transition region. The position of the
transition region strongly influence the computed \La
flux, which in their case is stronger than the observed one by a factor of 10.

{}
\section{The radiative losses}

\def\f{\matono {\Phi } \ifspace}
To estimate the energetic balance in the photosphere and chromosphere of
the star, it is important to know the radiative losses due to the
different species. Here we compute the radiative cooling rate \f
(ergs cm\u-3 \ s\u-1 ), \ie the net amount of
energy radiated at a given depth by the atmosphere, which is given by
$$ \f = 4 \pi \int \kappa\n\ (S\n-J\n)\ d\nu    \ .   \eqno(1)$$

In this study we computed the contributions due to \Hm, H, He I, Mg I and
II, Ca I and II, Fe I, Si I, Na I, Al I and CO. The overall results and the
most important individual contributions are presented in Fig. 9. A positive
value implies a net loss of energy (cooling), and a negative value represents
a net energy absorption (heating).

%\begfigwid 16 cm
%\begfigwid 1 cm
%\figure9{Total net radiative cooling rate for our model (full line), and
%the most important contributions to it}
%\endfig

It can be seen that in the region around the temperature minimum, \Hm is
a net heating agent. At this height, in fact, there is a missing {\it
cooling} agent, a fact that was already noted for the Sun (\eg
Mauas et al. 1990). However, it should be pointed out that
in this region CO is a strong coolant, and that the addition of the
other molecules present in this star should bring the model closer to
radiative balance.

The energy radiated above the temperature minimum gives an estimate of
the amount of chromospheric heating required to sustain this model. It
can be seen that there are two peaks, one in the mid-\-chromos\-phere and
the other in the transition region. A similar effect can be seen in the Sun
(see Vernazza et al. 1981).

\Hm is the main cooling agent in the low and mid\-chromos\-phere, while the
hydrogen bound-bound and bound-free transitions are important in the
high chromosphere
and in the transition region. We show in Fig. 10 a detailed analysis of the
cooling rates in this region of the atmosphere.
It can be seen that from 7000 to 15000K the
main contribution to the hydrogen cooling rate comes more or less equally from
\Ha,
the remaining Balmer lines, and the Lyman and Balmer continua. Above
15000K the cooling is almost exclusively due to \La.

%\begfig 8 cm
%\begfig 1 cm
%\figure{10}{Detail of the hydrogen cooling rates. The difference between
%the total rate (heavy line) and the hydrogen rate (full line), is due to
%Mg II. Bal refers to the Balmer lines from \Hb to H\sub8 , and continua
%are the Lyman and Balmer continua}
%\endfig

Another interesting feature is the fact that around 12000K the contribution
of the Mg II
lines becomes important, with a cooling comparable to that of \Ha.
In fact, the difference between the hydrogen rate and the total in Fig.
10 is due to Mg II.

It should be pointed out that in this paper we were not able to match the
observed flux in the Ca II K line, which may become an important component.
However, as this flux is maximum in the region were the
hydrogen emission takes place, we do not believe a more
accurate value for this flux would change the global values presented here.

Conversely, our computed flux for \La is about 3 times larger
than the observations. This fact should be kept in mind when considering
the rates above 15000K, where \La is the main cooling agent.

\section{Conclusions}

We present a semi-empirical atmospheric model for the dMe star AD Leo.
This constitutes the first model computed to match a wide set
of observations, from the continuum emission to a set of several
chromospheric lines. The agreement found between the observed and
computed spectral features is generally good.

In particular, the continuum calculations presented here match quite
well the observations when line blanketing is included. On the other
hand, we show that the calculations done neglecting line blanketing grossly
overestimates the continuum emission. This indicates that models based
on continuum observations computed without properly including line
blanketing should be taken with caution.

The main disagreement between the observations and the computed profiles
reside in the Ca II K line. In fact, while the calculations for \Ha and \La
overestimate the observations, the computed intensity for Ca II K is
almost a factor of 4 smaller than observed, and we were not able
to get a better agreement for Ca II K without completely destroying the
agreement for the other lines. A similar problem was found by Giampapa et al.
(1982), who concluded that it is not possible with a homogeneous model to fit
both the Ca and the Mg II h and k lines. Similarly, Hawley \& Fisher (1992)
had problems matching the Ca and Mg II fluxes.

This suggests that the problem may be due not to our chromospheric model,
but to an incorrect treatment of the atomic parameters of Ca II, in
particular to the not inclusion of collisions with hydrogen.

Finally, we would like to point out that there are some uncertainties in
the atmospheric model, mainly due to the fact that no reliable
diagnostics of the \Tm region is available. To find an indicator of the
temperature in this region, observations with better spectral resolution are
needed.
\bigskip
{\it Acknowledgements} We would like to thank Dr. S. Hawley for providing us
with the
computer files with the observations, and Dr. R. Falciani for very
helpful discussions. We would also like to thank Dr. E. Avrett for the
Pandora computer program, and F. Tribioli and R. Baglioni for their
patience managing the computer system. PM whishes to acknowledge
Comission 38 of the I.A.U., for financial help with his travel to
Arcetri.

\vfill
\eject
{}
\def\ref{\leftskip = 24pt \advance\leftskip by 24pt \parindent=-24pt}
\def\refindent{\advance\leftskip by 24pt \parindent=-24pt}
\heading{REFERENCES}

%\refindent Ambruster, C.W., Pettersen, B.R., and Sundland, S.R., 1989, A\&A
\ref Ambruster, C.W., Pettersen, B.R., and Sundland, S.R., 1989, A\&A
208, 198.\par

%\refindent \Avr, \Loe, 1992, in: The 7th. Cambridge Workshop on Cools Stars,
\ref \Avr, \Loe, 1992, in: The 7th. Cambridge Workshop on Cools Stars,
	Stellar Systems, and the Sun, eds. M. Giampapa and J.A. Bookbinder \par

\ref Cram L.E., Mullan D.J,. 1979, ApJ  234, 579

\ref Doyle, J.G. 1987, MNRAS 224, 1P

\ref Giampapa M.S., Linsky, J.L., Schneeberger T.J., Worden S.P., 1978, ApJ
226, 144

\ref Giampapa M.S., Worden S.P., Linsky, J.L., 1982, ApJ  258, 740

\ref Hawley S.L., 1989, Ph. D. Thesis, The University of Texas at Austin. \par

\ref Hawley S.L., Fisher G.H., 1992, ApJS  78, 565 \par

\ref \Kur, 1991a, in: Stellar Atmospheres: Beyond classical models,
	eds. L. Crivellari, I. Hubeny, and D.G. Hummer, Kluwer, Dordrecht \par

\ref \Kur, 1991b, in: Precision Photometry: Astrophysics of the Galaxy,
	eds. A.G. Davis Philip, A.R. Upgren, and K.A. Janes, L. Davis Press,
	Schenectady \par
\ref Kanlakys G., 1985, J. Phys. B 18, L167 \par

\ref Kunkel W.E., 1975, in: IAU Symposium 67: Variable Stars and
	Stellar Evolution, eds. V.E. Sherwood and L. Plant, Reidel, Dordrecht \par

\ref Lites, B.W., and Skumanich, A. 1982, ApJS 49, 293 \par

\ref Mathioudakis, M., and Doyle, J.G., 1991, A\&A 244, 433 \par

\ref \Mau, \Avr, \Loe, 1988, ApJ 330, 1008 \par

\ref \Mau, \Avr, \Loe, 1989, ApJ 345, 1104 \par

\ref \Mau, \Avr, \Loe, 1990, ApJ 357 279. \par

\ref Metcalf, T.R, Canfield, R.C., \Avr, and Metcalf, F.T., 1990, ApJ
	350, 463. \par

\ref Mould J.R., 1976, A\&A 48, 443 \par

\ref Pettersen B.R., 1976, Catalogue of Flare Star Data, University of
	Oslo Rep. 46. \par

\ref Pettersen B.R., 1980, A\&A 82, 53 \par

\ref Pettersen B.R., Coleman L.A., 1981, ApJ  251, 571 \par

\ref Pettersen B.R., Coleman L.A., Evans, D.S., 1984, ApJS 54, 375 \par

\ref Pettersen B.R., Hawley S.L., 1987, Institute of
	Theoretical Astrophysics Publication Series No. 2 \par

\ref Pettersen B.R., Hawley S.L., 1989, A\&A 217, 187 \par

\ref Pettersen B.R., Panov K.P., Sandman W.H., Ivanova M.S., 1986, A\&AS,
	66, 235 \par

\ref Pettersen B.R., et al. 1990, in: Flare Stars in: Star Clusters,
	 Associations and the Solar Vicinity, eds. L.V. Mirzoyan et al.
	 Kluwer, Dordrecht, p. 15 \par

\ref Saar S.H., 1990, in: Solar Photosphere: Structure, Convection, and
	 Magnetic Fields, ed. J.O. Stenflo p. 427 \par

\ref Saar S.H., Linsky J.L., 1985, ApJ 299, L47 \par

\ref Schaifers, K., and Voigt, H.H., 1982, Numerical Data and Functional
	Relationships in Science and Technology, vol. 2, p. 52. \par

\ref Spiesman, W.J., and Hawley, S.L., 1986, ApJ 92, 664. \par

\ref Sundland S.R., Pettersen B.R., Hawley S.L., Kjeldseth-Moe O.,
	Andersen B.N., 1988, in: Activity in cool stellar envelopes, eds. O.
	 Havnes et al., Kluwer, Dordrecht \par

\ref Vernazza, J.E., \Avr, \Loe, ApJS 45, 635. \par

\ref Vogt S.S., Soderblom D.R., Penrod G.D., 1983, ApJ  269, 250. \par

\ref Worden S.P., Schneeberger T.J., Giampapa M.S., 1981, ApJS 46, 159 \par
%\endref
\vfill
\eject

{\bf Figure captions}

Figure 1 {Temperature distribution for our atmospheric model.
Also shown are the heights of formation of the lines used in the modeling.
The dashed line represents the model by Hawley \& Fisher (1992)}

Figure 2 {Comparison of the observations (full line) in the continuum with the
results computed for our model, with (dashed line) and whitout (dotted
line) considering line blanketing. Also indicated are the heads of the most
important molecular bands}

Figure 3 {Comparison of the observations (crosses) in the Balmer lines and the
results computed for our model. Here and in the next figures, the dashed
line shows the computed profile with ``infinite" spectral resolution, and the
full line shows the calculations smeared to the same resolution than
the observations}

Figure 4 {Planck function ($B_\nu$, full line) and source function
($S_\nu$, dashed line) for different lines. The depth of formation (peak
of the contribution function) of the line center is also shown. The
values at disk center are indicated above the $S_\nu$ curve, and those
at $\mu=0.6$ are indicated below it}

Figure 5 {Comparison of the observations in the Na D and infrared  lines and
the
results computed for our model. See fig. 3}

Figure 6 {Comparison of the observations in the Mg b lines and the
results computed for our model. See Fig. 3}

Figure 7 {Comparison of the observations in the Ca I \lin4227 in the quiet
spectrum (solid line) and the flare spectrum at two different times (dotted
line)}

Figure 8 {Comparison of the observations in the Ca II K and infrared line and
the
results computed for our model. See Fig. 3}

Figure 9 {Total net radiative cooling rate for our model (full line), and
the most important contributions to it}

Figure {10} {Detail of the hydrogen cooling rates. The difference between
the total rate (heavy line) and the hydrogen rate (full line), is due to
Mg II. Bal refers to the Balmer lines from \Hb to H\sub8 , and continua
are the Lyman and Balmer continua}

\bye

\section{ INTRODUZIONE}

Tre anni fa era circolato un fac simile da adottare per la preparazione
degli Arcetri Astrophysics Preprints (Bandiera \etal\ 1987).
Nel frattempo molte cose sono cambiate.
Innanzitutto ormai pochi autori usano MacWrite (o programmi similari)
su MacIntosh per la produzione dei testi; invece programmi come TeX o
LaTeX (operativi su SUN) sono sempre pi\`u diffusi.
Lo stesso ``supporto cartaceo'', passando allo standard A4, ha
richiesto un aggiornamento nel formato della pagina.

Inoltre, in questi ultimi tempi diverse importanti riviste hanno
sentito la necessit\`a di ridefinire i loro standard, allo scopo di
uniformarli secondo le direttive della Commissione 5 dell'IAU
(Abt 1990).
Files di macro TeX sono gi\`a disponibili per la formattazione di
testi secondo i suddetti standard, almeno per The Astrophysical Journal
e per Astronomy \&\ Astrophysics.
Nel nostro piccolo, anche noi cerchiamo di adeguare il nostro
standard, possibilmente in compatibilit\`a con quelli delle riviste
maggiori.

\section{ IL NUOVO STANDARD AAP}

\firstsubsection{Formato Pagina}

Il formato deve essere compatibile con fogli formato A4, che andranno
spillati assieme sulla costola.
La stampa (testo, figure, note a pi\'e di pagina, e numerazione pagina)
deve essere confinata in una regione larga non pi\`u di 16.5 cm, ed
alta non pi\`u di 25 cm.
Valori ragionevoli sono quelli standard di TeX (6.5 in per 8.9 in).

Per i caratteri si usino le dimensioni normali dei caratteri di macchina
da scrivere, con uno spazio interlinea di circa 6 mm.
Insomma, i caratteri non devono apparire troppo piccoli, ma le righe
nemmeno troppo spaziate (come in alcune copie per referee).
In TeX tutto ci\`o si ottiene col comando
{\tt \char'134magnification=\char'134magstep1}.

A titolo preventivo, si avverte che non saranno mai accettati n\'e
preprints con testo a due colonne, n\'e in formato ``landscape''.
Anche per Figure e Tabelle, ove possibile, il basso dovr\`a essere
orientato verso il basso!

\subsection{Frontespizio}

Nello spazio corrispondente alla finestrella della copertina vanno
racchiusi il titolo dell'articolo, i nomi degli autori (con indici
numerici ad indicare le affiliazioni), ed il codice del preprint
stesso (costituito da codice progressivo ed ultime due cifre dell'anno).
Verso met\`a pagina sono listate le affiliazioni (istituti di appartenenza
con indirizzo), e similari (tipo {\sl Visiting Astronomer\/}, {\sl
Visiting Associate\/,} etc.).

Verso il basso invece deve apparire il nome della rivista, seguita dal
classico {\sl in press}.
Una dicitura alternativa (pi\`u consona per Proceedings) \`e {\sl To
appear in}.
Per lavori invitati a congressi si usi invece {\sl Invited talk\/} (o
{\sl review\/}) {\sl at\/} seguito da titolo del congresso, data e luogo
del medesimo.
Si ricorda che di norma gli AAP si compongono solo di articoli gi\`a
accettati per pubblicazione su riviste, o testi di discorsi invitati
che appariranno su Proceedings.

\subsection{Pagine prima del testo}

La prima pagina, come si \`e detto, \`e il frontespizio.
La pagina contenente gli indirizzi degli autori \`e stata assimilata
nel frontespizio.
Segue una pagina contenente l'Abstract; se l'Abstract \`e sufficientemente
corto, conviene lasciare in alto un p\'o di spazio (per evitare che
appaia in trasparenza dietro il titolo).
Nei rari casi in cui l'Abstract superi una pagina di lunghezza, si
cambi pagina prima di iniziare il testo vero e proprio.
Tutte le pagine descritte finora non vanno numerate.

\subsection{Numerazione Pagine}

La numerazione inizia col testo vero e proprio, e parte col numero 1
(sorpresi?).
Si suggerisce, secondo lo standard TeX, di numerare le pagine al centro
in basso.
La numerazione deve proseguire con lo stesso formato anche sulle
Tavole nonch\'e sulle Figure.

\subsection{Tavole e Figure}

Tavole e Figure devono essere contraddistinte da numerazione
progressiva.
Tavole che si estendono su pi\`u di una pagina devono essere riferite
su ognuna (magari con l'aggiunta di un {\sl
continue\/}, nelle successive).
Ricordando quanto gi\`a detto nella sezione sul Formato Pagina, anche
Tavole e Figure devono rientrare (compresa la numerazione) nelle
dimensioni massime prescritte.

\subsection{Impaginazione}

Si segua in genere le prescrizioni della rivista.
A puro titolo di riferimento, dopo il testo ci aspetteremmo eventuali
Appendici, Tabelle, Referenze, Figure Captions e relative Figure.

Per rispetto dell'origine ligure del responsabile, nonch\'e
dell'integrit\`a della Foresta Amazzonica, si raccomanda di
utilizzare lo spazio con parsimonia, non cambiando pagina ad ogni
nuova sezione, salvo quando lo spazio rimasto sia poco, e raggruppando
in una stessa pagina pi\`u Figure o Tavole, se di piccole dimensioni.
Ovviamente la parsimonia non vada a contrastare col senso estetico.

\section{ COMPITI DEGLI AUTORI}

Il lavoro di tutti (segretarie, addetti alla tipografia, e finanche
utenti) pu\`o essere snellito considerevolmente se gli autori seguono
alcune semplici norme prima e durante la preparazione dei preprints.

\subsection{Che fare}

Allorch\'e l'articolo \`e stato accettato dalla rivista, ed il testo
del preprint \`e pronto, occorre informare gli uffici segreteria primo
piano, indicando titolo ed autori del lavoro, la rivista su cui
apparir\`a, nonch\'e il numero copie personali richiesto.
Si consiglia di non eccedere nel numero richiesto, visto che gli AAP
vengono gi\`a automaticamente inviati ai maggiori istituti astronomici,
e che un preprint perde ben presto la propria utilit\`a, con
la pubblicazione dell'articolo sulla rivista.
La segreteria assegner\`a ad ogni preprint un numero d'ordine, che
dovr\`a essere riportato sul frontespizio.
I lavori necessari alla sistemazione dell'impaginazione sul formato
AAP sono a carico degli autori, che si accorderanno personalmente con
le segretarie per commissionare parte del lavoro.

Dopo variazioni eseguite da terzi, gli autori sono responsabili di
eventuali errori apportati, e quindi sono tenuti a controllare
accuratamente la versione finale, prima della stampa.
A titolo di suggerimento, ricettacoli di errori possono essere non solo
la numerazione e l'ordine delle tabelle, delle figure, i nomi o gli
indirizzi degli autori, ma (ed \`e successo) perfino lo stesso titolo
dell'articolo!

Il testo va consegnato, completo in ogni sua parte, al responsabile o
agli addetti al servizio tipografia, i quali verificheranno se il
preprint \`e in regola per la pubblicazione.
In caso di dubbi durante la sistemazione si raccomanda di contattare
i medesimi preventivamente, per discutere assieme le variazioni da
apportare.

\subsection{Che fare prima}

Capita spesso che autori siano in possesso di testi che non possono
essere riformattati, in quanto prodotti altrove, o di Tabelle e Figure
scarsamente decifrabili, perch\'e fotocopie di fotocopie di fotocopie.
Per evitare, o quantomeno ridurre tali inconvenienti si raccomandano
alcuni piccoli accorgimenti.

Innanzitutto, prima di spedire il plico alla rivista occorre
assicurarsi di essere in possesso di copie di buona qualit\`a del testo,
delle Tavole, e soprattutto delle Figure.
Quanto al testo, meglio sarebbe averne una copia ``digitale'', magari
facendosela inviare dai co-autori attraverso posta elettronica.

\section{ CONCLUSIONI}

Si ringrazia per l'attenzione e per la futura collaborazione.
A titolo di ``gentile presente'', si informa che il file utilizzato
per la stesura del presente testo \`e disponibile sulla directory
SUN {\tt /usr/tex/inputs} sotto il nome {\tt aap\_facs.tex}.
Agli utenti TeX si consiglia di usare tale file come template per la
compilazione dei documenti, ed in particolar modo per la redazione
del frontespizio.

\heading{ RINGRAZIAMENTI}

Siamo grati al Dr.\ J.~van Hardy per interessanti discussioni
sull'argomento, un'attenta lettura del manoscritto, ed utili
commenti critici.

\def\refindent{\advance\leftskip by 24pt \parindent=-24pt}
\def\journal#1#2#3#4#5{{\refindent{#1} {#2}, {#3}, {#4}, {#5}.\par}}
\heading{ REFERENZE}
\journal{Abt, H.A.}{1990}{ApJ}{357}{1}

\refindent Bandiera, R., Basile, P., Felli, M., e Venturi, R. 1987,
  AAP n.\ 0/87.\par

\bye